\documentclass[twocolumn,showpacs,preprintnumbers,amsmath,amssymb,aps,prb,longbibliography]{revtex4-2} 
\usepackage{graphicx}
\usepackage{dcolumn}
\usepackage{bm}
\usepackage{verbatim} 
\usepackage[usenames, dvipsnames]{color} 
\usepackage{ulem}
\usepackage{color}
\usepackage[colorlinks=false,pdfencoding=auto,psdextra]{hyperref}
\usepackage{multirow}
\usepackage{tabularx}
\usepackage{dsfont}

\definecolor{orange}{rgb}{1.0, 0.5, 0.0}

\begin{document}

\title{Continuous time crystal in an electron-nuclear spin system: \\ stability and melting of periodic auto-oscillations}


\author{A.~Greilich$^{1}$, N.~E.~Kopteva$^{1}$, A.~N.~Kamenskii$^{1}$, P.~S.~Sokolov$^{1}$, V.~L.~Korenev$^{2}$, M.~Bayer$^{1}$}

\affiliation{$^{1}$Experimentelle Physik 2, Technische Universit\"at Dortmund, 44227 Dortmund, Germany}
\affiliation{$^{2}$Ioffe Institute, 194021 St. Petersburg, Russia}


\maketitle

\textbf{Crystals spontaneously break the continuous translation symmetry in space, despite the invariance of the underlying energy function. This has triggered suggestions of time crystals analogously lifting translational invariance in time. Originally suggested for closed thermodynamic systems in equilibrium~\cite{WilczekPRL12,ShaperePRL12}, no-go theorems prevent the existence of time crystals~\cite{Nozieres_2013,BrunoPRL13,WatanabePRL15}. Proposals for open systems out of equilibrium led to the observation of discrete time crystals~\cite{ZhangNat17,ChoiNat17} subject to external periodic driving to which they respond with a sub-harmonic response.~\cite{IeminiPRL18,BucaNatComm19,KesslerPRA19} A continuous time crystal is an autonomous system that develops periodic auto-oscillations when exposed to a continuous, time-independent driving ~\cite{IeminiPRL18,BucaNatComm19,KesslerPRA19}, as recently demonstrated for the density in an atomic Bose-Einstein condensate with a crystal lifetime of a few ms~\cite{KongScience22}. Here we demonstrate an ultra-robust continuous time crystal in the nonlinear electron-nuclear spin system of a tailored semiconductor with a coherence time exceeding hours. Varying the experimental parameters reveals huge stability ranges of this time crystal, but allows one also to enter chaotic regimes, where aperiodic behavior appears corresponding to melting of the crystal. This novel phase of matter opens the possibility to study systems with nonlinear interactions in an unprecedented way.}

The spins of electrons and nuclei in semiconductors form a system showing strongly intertwined, highly nonlinear dynamics. Several decades ago, two fundamental problems that have remained unresolved so far were formulated for this system. The first one is the spontaneous phase transition to a magnetically ordered, anti-ferromagnetic nuclear state canceling the fluctuations in the nuclear spin ensemble. This state is expected to develop by deep cooling of the nuclei to temperatures in the order of $10^{-7}$\,K~\cite{Merkulov1998Polaron,OultonPRL2007,ScalbertPRB17,FischerPRB20,VladimirovaPRB21,KleinjohannPRB22}, 
Recently, advanced cooling protocols allowed achieving record nuclear spin temperatures of $5\times 10^{-7}$\,K in GaAs~\cite{Kotur21}, which, however, was not yet sufficient for the phase transition. 

The second problem is the existence of a strange attractor in the chaotic polarization oscillations of the autonomous electron-nuclear spin system (ENSS)~\cite{OptOR_Flei_Merk}. Depending on the number of accessible oscillation frequencies, the ENSS can have different types of attractors in phase space. A periodic auto-oscillation of one frequency is called a limit cycle, while an auto-oscillation with incommensurate frequencies occurs on a two-dimensional torus. 
Indications of these trajectories were found by Kalevich~\textit{et al.}~\cite{Kalevich1993}. Further theoretical work also predicted the existence of homoclinic trajectories~\cite{Bakaleinikov1994} that do not form closed cycles returning to their start positions - a precursor of chaotic motion. Up to now, this range of chaos has not been observed yet in semiconductors due to unrealistic required parameters, including the Overhauser effective magnetic field of  polarized nuclei.


\begin{figure*}[t!]
\begin{center}
\includegraphics[width=16cm]{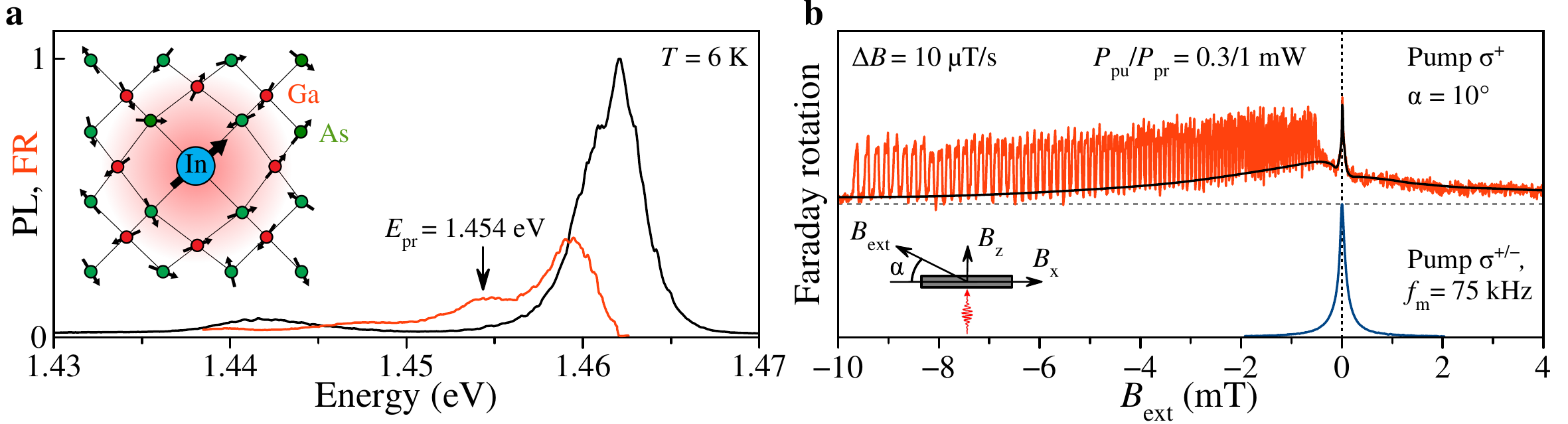}
\caption{\label{fig1} \textbf{Optical properties of the Si-doped In$_{0.03}$Ga$_{0.97}$As epilayer.}
\textbf{a}, Photoluminescence spectrum at $T = 6$\,K, excited by a diode laser at $E_\text{pu} = 1.579$\,eV photon energy, the black curve. The red curve gives the Faraday rotation spectrum, measured by scanning a probe laser. For electron spin polarization, the same pump laser was used for recording the photoluminescence. The inset shows a sketch of the GaAs lattice distortion by a large In atom replacing a Ga atom.
\textbf{b}, Faraday rotation measured at the probe energy $E_\text{pr} = 1.454$\,eV while scanning the tilted magnetic field, using pump excitation with helicity modulation at $f_\text{m}=75$\,kHz frequency (blue curve) and with fixed helicity (red curve). The field sweep rate is $\Delta B = 10$\,$\mu$T/s, the tilt angle between the sample plane and the magnetic field is $\alpha = 10^\circ$ (see sketch at the bottom). The black curve shows the FR nominally expected for circular excitation, to highlight the additional maximum at the finite magnetic field strength of about $-1$\,mT due to nuclear polarization. This curve is gained from the red curve by removing the oscillations. The shape of curves is independent of the field scan direction.}
\end{center}
\end{figure*}

Here, we use a semiconductor with an ENSS tailored to develop pronounced auto-oscillations under continuous optical pumping with circularly polarized laser light. The ENSS with an on-purpose reduced symmetry is an In$_{0.03}$Ga$_{0.97}$As epilayer doped with Si donors. The observed, strictly periodic auto-oscillations are unique signatures of a continuous time crystal (CTC) with an exceptionally long-lasting coherence exceeding hours, limited only by the measurement time, or in other words, a near-perfect clocking order of the "time atoms". The robustness of this first solid-state CTC is tested by linear and nonlinear analysis tools and approved across order-of-magnitude wide ranges of control parameters (laser power, sample temperature, and magnetic field), in which limit cycles in phase space are found for the ENSS evolution. On the other hand, these parameters allow one to vary the CTC period. Further, we could enter for the first time external parameter ranges of chaotic behavior, evidenced by a fractional correlation dimension~\cite{Grassberger1983}, a positive maximum Lyapunov exponent, and a $K$-parameter value close to unity in the $0-1$ chaos test~\cite{Gottwald01test,Toker2020}. Then the chaotic auto-oscillations violate the ideal periodicity in time and can be interpreted as CTC “melting”.

\section*{Results}

\subsection*{Optical properties of the epilayer}

In our studies, the sample was tailored for a reduction of the crystal symmetry from cubic by introducing locally well-defined lattice distortions that impact the ENSS through the nuclear level splitting. The result of these considerations was an In$_{0.03}$Ga$_{0.97}$As epilayer doped with Si donors whose optical properties are shown in Fig.~\ref{fig1}a. Localization of electrons at the Si-donors leads to an enhanced hyperfine interaction with the nuclei, compared to free electrons. 

At the temperature of $T = 6$\,K, the dominant spectral line in the photoluminescence (PL) spectrum is associated with the recombination of free excitons, while the emission at its low energy flank arises from donor-bound exciton recombination; see the black curve. Due to the incorporated indium atoms, the reduced band gap of the epilayer at $E_g$(InGaAs) $ = 1.461$\,eV is located in the transparency range of the GaAs substrate with $E_g$(GaAs) $ = 1.519$\,eV. Spin polarization of the Si-bound electrons is generated by non-resonant circularly polarized excitation, using a continuous wave (CW) pump diode laser emitting at $E_\text{pu}=1.579\,$eV photon energy. The pump-induced spin polarization is monitored through the Faraday rotation (FR) of the linear polarization of the probe beam emitted by a tunable CW laser. The red curve in Fig.~\ref{fig1}a shows the spectral dependence of the FR signal recorded by tuning the probe energy. For all further experiments, we fix the energy of the probe laser at $E_\text{pr}=1.454\,$eV corresponding to a local FR signal maximum, see the arrow in Fig.~\ref{fig1}a. This choice is thus simultaneously optimized for low laser absorption and large FR. More details are given in the Methods.

\begin{figure*}[t!]
\begin{center}
\includegraphics[width=16cm]{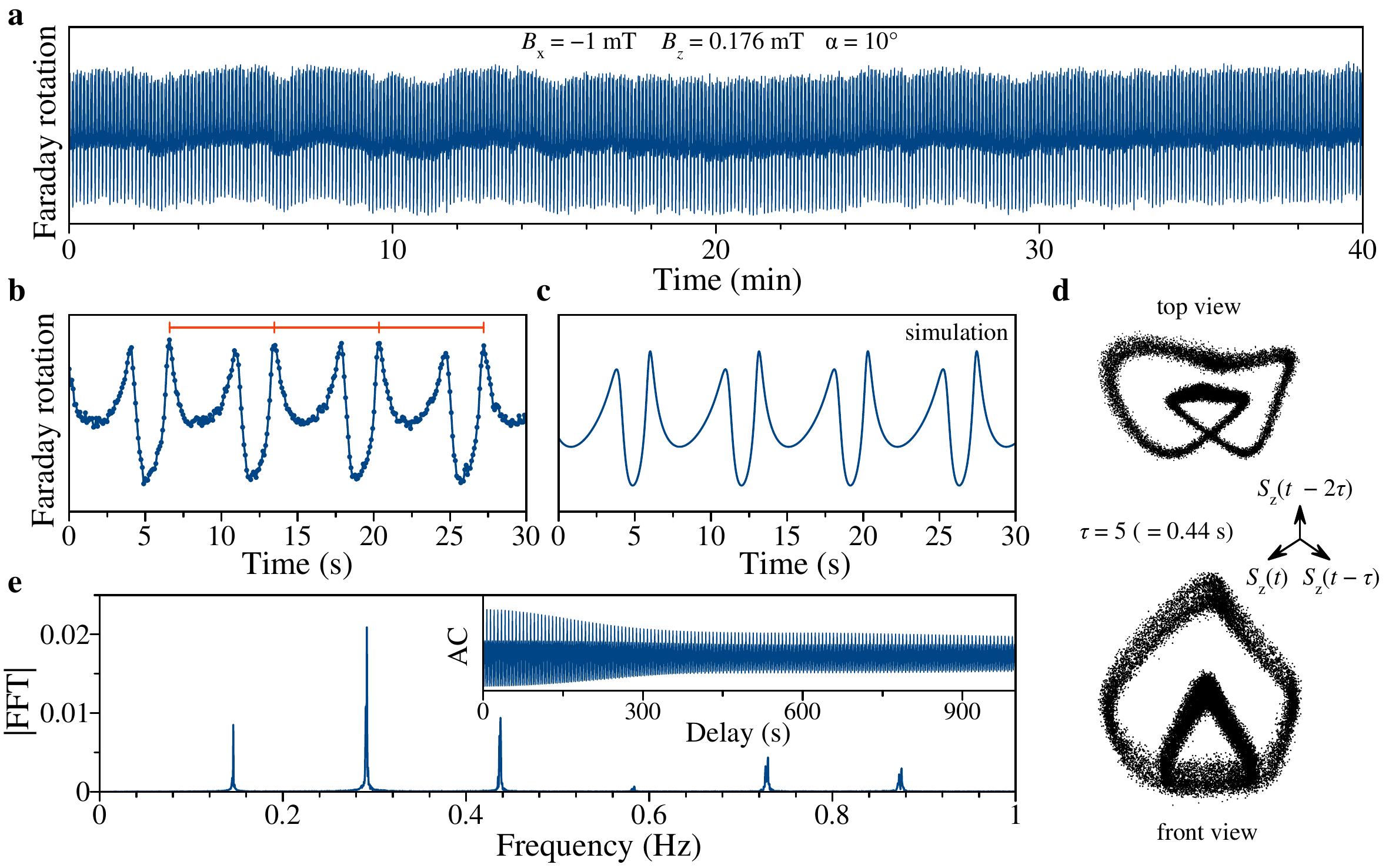}
\caption{\label{fig2} \textbf{Periodic auto-oscillations of CTC.}
\textbf{a}, Oscillations of the electron spin polarization in time monitored in FR, applying a tilted magnetic field with components $B_x = -1$\,mT and $B_z = 0.176$\,mT, corresponding to $\alpha=10^{\circ}$. Pump and probe photon energies: $E_\text{pu} = 1.579$\,eV, $E_\text{pr} = 1.454$\,eV; pump and probe powers: $P_\text{pu} = 0.3$\,mW, $P_\text{pr} = 1$\,mW.
\textbf{b}, Initial 30 seconds of the temporal data range from panel \textbf{a}, zooming into the oscillation details. The set of red segments of equal length between the FR maxima highlights the CTC periodicity.
\textbf{c}, Calculated electron spin polarization, using the parameters $\alpha = 10^\circ$, $B_x = -1$\,mT, $a_\text{N} = 20$\,mT, $b_\text{N} = 21$\,mT, and $T_\text{N} = 0.5$\,s (see the Interpretation section for details).
\textbf{d}, Top and front view of the three-dimensional plot of the spin polarization cycle, with the coordinates successively delayed by $\tau \delta t=5\delta t$, where the measurement time step $\delta t = 89\,$ms. Such a delayed coordinates choice allows the representation of the time series in three dimensions. The black points mark the data.
\textbf{e}, Fast Fourier transform and autocorrelation function as a function of delay time (inset), calculated for the signal from panel \textbf{a}. 
}
\end{center}
\end{figure*}

The best way to assess the interaction between the electron and nuclear spins is to monitor the electron spin polarization, first done as a function of a tilted magnetic field $B_\text{ext}$, see the sketch at the bottom of Fig.~\ref{fig1}b indicating the field orientation in combination with its components normal and parallel to the sample plane. Tilting the magnetic field is favorable for building up nuclear spin polarization and measuring electron spin dynamics:
The initial electron spin orientation is set by the helicity of the circularly polarized pump laser, which orients the spins along the beam direction ($z$-axis)~\cite{OptOr_book}. Subsequently, the electron spins precess about the transverse component $B_\text{x}$ of the magnetic field and lose their initial orientation. The pump-induced electron polarization can be transferred to the nuclear spins by flip-flop processes~\cite{OptOR_Flei_Merk}. The transfer efficiency increases when a magnetic field is applied along the electron spins, i.e. by the $B_\text{z}$ component. However, when the pump light helicity is modulated sufficiently fast between right and left circular polarization, the available time for nuclear polarization build-up is too short, which is the case when the modulation frequency is higher than the rate of nuclear polarization by the electron spins. Then the observed signal has the Lorentzian shape called the Hanle curve~\cite{Hanle} with a width inversely proportional to the electron spin lifetime, see the blue curve centered at zero magnetic field in Fig.~\ref{fig1}b. Here, the half width at half maximum of the peak is 0.1\,mT, which corresponds to the electron spin lifetime $T_s = 200\,$ns, using the electron $g$-factor $-0.568$~\cite{RittmannPRB2022}. 

In contrast, if the excitation is done with fixed circular polarization, the electron spins polarize the nuclear spin system, which produces an effective magnetic field, the Overhauser field, acting back on the electron spins and changing the FR response to the external magnetic field. In a tilted magnetic field, the curve becomes strongly asymmetric relative to zero field and demonstrates a FR maximum shifted to finite $B_x$ by the Overhauser field strength. For clarity of this expected behavior, we show the black curve in Fig.~\ref{fig1}b, which is a spline fit to the original data given by the red curve that removes the superimposed oscillations. Here, one nicely sees the side peak at about $B_x = -1$~mT. Surprisingly, at the chosen slow sweep rate of the field, $\Delta B= 10\,\mu$T/s, in addition, strong polarization oscillations emerge at negative fields. The appearance of these oscillations depends on the field sweep rate, and for rates higher than the chosen one, the curve develops into the black curve (not shown here).

\subsection*{Periodic auto-oscillations}

The observation central to this manuscript is the auto-oscillations of the ENSS: when keeping the experimental parameters fixed in the right ranges, see below, the temporal evolution of the FR signal is strictly periodic without any notable decay, as demonstrated in Fig.~\ref{fig2}a across the whole time range in a geometry with $B_x = -1$\,mT and $B_z = 0.176$\,mT ($\alpha=10^\circ$): Amplitude and frequency of the oscillations remain constant over 40 minutes measurement time, only the background varies slightly. Figure~\ref{fig2}b gives a close-up of the FR signal with an M-shape repeating with a period of about 6.9\,s. To analyze the signal, we calculate its fast Fourier transformation (FFT), shown in Fig.~\ref{fig2}e. The spectrum consists of a distinct set of equidistantly spaced spikes,  corresponding to the precession frequency of 0.145\,Hz and its higher harmonics. The spikes in the frequency comb are remarkably narrow with a width of 0.4\,mHz, as expected from the decay-free FR trace, limited only by the signal accumulation time of 40 minutes. The inset in Fig.~\ref{fig2}e shows the autocorrelation (AC) function of the trace, showing a correlation between the signal with a delayed copy of itself. Also here, the AC amplitude does not decay with increasing delay, corroborating the non-random periodicity of the signal. 

\begin{figure*}[t!]
\begin{center}
\includegraphics[width = 18cm]{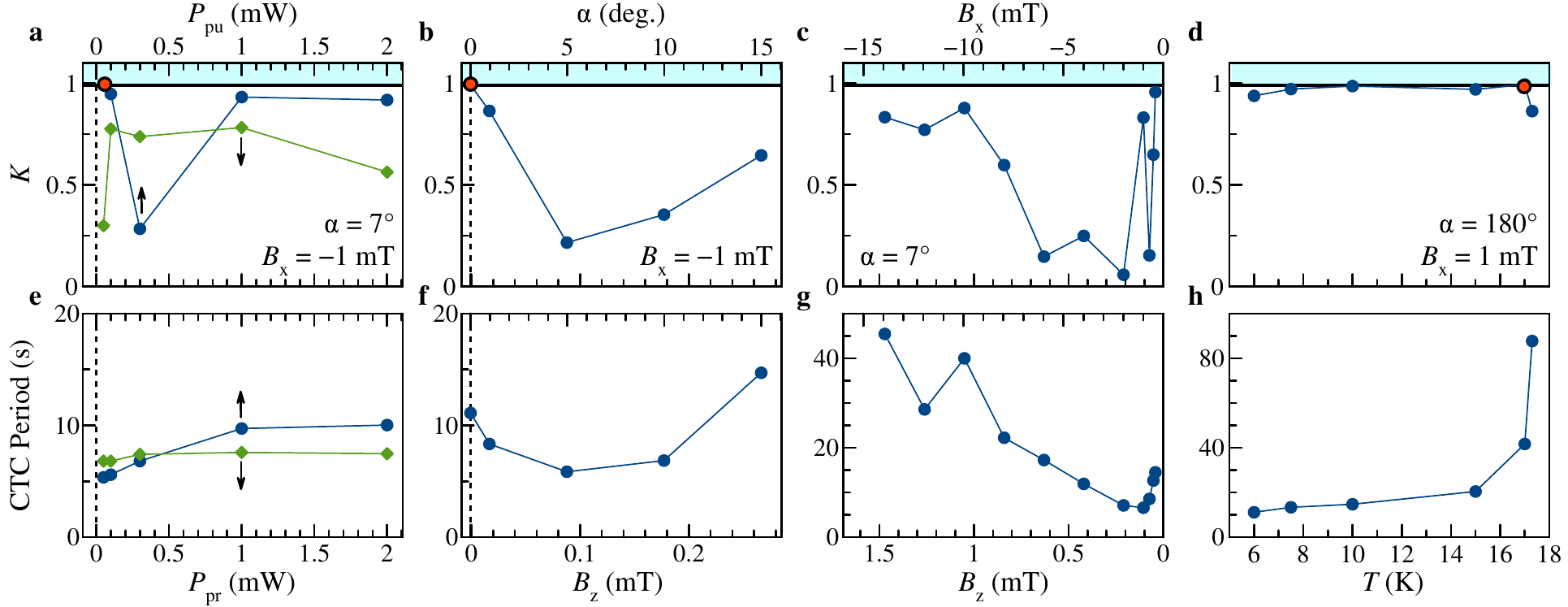}
\caption{\label{fig3} \textbf{Dependencies of the $K$-parameter and CTC period on the experimental conditions.}
The top (bottom) axis of the upper (lower) figure is also relevant for the other figure in the column. 
\textbf{a}, The green diamonds show the dependence of the $K$-parameter on the probe power at $P_\text{pu} = 0.3$\,mW (lower axis), and the blue circles give the dependence on the pump power at $P_\text{pr} = 0.05$\,mW (upper axis). $B_x = -1$\,mT, $\alpha = 7^\circ$, and $T=6$\,K. 
\textbf{b}, $K$ for different $B_z$ at fixed $B_x = -1$\,mT. The upper axis shows the corresponding angle $\alpha$. $P_\text{pu} = 0.3$\,mW and $P_\text{pr} = 1$\,mW, $T=6$\,K.
\textbf{c}, $K$ dependence on the strength of the magnetic field at fixed $\alpha = 7^\circ$. The magnetic field's $B_z$ and $B_x$ components are plotted on the lower and upper axes, respectively. $P_\text{pu} = 0.3$\,mW and $P_\text{pr} = 0.05$\,mW, $T=6$\,K.
\textbf{d}, $K$ dependence on temperature. $B_x = 1$\,mT at $\alpha = 180^\circ$. $P_\text{pu} = 0.3$\,mW and $P_\text{pr} = 1$\,mW.
In all cases, the red-colored circles mark the signal with $K>0.99$.
\textbf{e}, The green diamonds show the dependence of the CTC period on the probe power at $P_\text{pu} = 0.3$\,mW (lower axis), and the blue circles give the dependence on the pump power at $P_\text{pr} = 0.05$\,mW (upper axis).
\textbf{f}, CTC period for different $B_z$ at fixed $B_x = -1$\,mT.
\textbf{g}, CTC period dependence on the strength of the magnetic field at fixed $\alpha = 7^\circ$.
\textbf{h}, CTC period dependence on temperature.}
\end{center}
\end{figure*}

Auto-oscillations occur in dissipative, nonlinear systems as the ENSS, when a continuous source of incoming energy - in our case applied highly non-resonant - compensates for the energy losses. The time evolution is then determined by the intrinsic properties of the system and is not affected from the outside. It may evolve as periodic, chaotic, or even unpredictable. We find clear indications for periodicity in our case, allowing a reliable claim for CTC behavior. As we observed no indication of any oscillation decay for 40 minutes, we can safely conclude a CTC lifetime of at least a few hours. The FFT of the oscillations can be understood as a structural analysis of the CTC, characteristic for the crystal unit cell.

We have performed additional crucial tests established in the chaos theory of nonlinear systems. Application of these tests becomes possible only because of the long lifetime of the CTC. One such time-series analysis test is the Chaos Decision Tree Algorithm described in Ref.~\cite{Toker2020}. The analysis gives the parameter $K$ in the $0-1$ chaos test, which approaches zero for periodic systems, while it converges to unity for chaotic systems. For the data presented in Fig.~\ref{fig2}, $K=0.3536$, which approves periodicity. Additional supporting measures are the correlation dimension ($D_2$)~\cite{Grassberger1983} and the maximal Lyapunov exponent (LE)~\cite{ROSENSTEIN1993117}. In our case, they are calculated using the TISEAN software package~\cite{Hegger1999}. If the correlation dimension is an integer, the time series describes a periodic signal, and the maximal Lyapunov exponent should be less than or equal to zero. In our case, the periodicity is supported by the correlation dimension $D_2=1.1\pm 0.1$ and the non-positive maximal Lyapunov exponent (see also Extended Data Fig.~\ref{figS6} for the full analysis).

A useful way to visualize the time evolution of a nonlinear system is to use delayed coordinates. A spin polarization vector is constructed so that the coordinates are chosen with a particular time delay $\tau$ relative to each other. The coordinates for each phase space point are calculated as $S(x,y,z) = (S(t), S(t-\tau), S(t-2\tau))$, with $S(t)$ being the spin polarization at time $t$. Figure~\ref{fig2}d shows two views of this spin polarization vector using the time step of $\tau\delta t= 5\delta t$, with $\delta t=89\,$ms. Clearly, the phase space trace is a limit cycle as required for the periodic oscillations of a CTC. Next, we check the stability of the CTC.

\subsection*{Stability and melting of CTC}

\begin{figure*}[t!]
\begin{center}
\includegraphics[width=16cm]{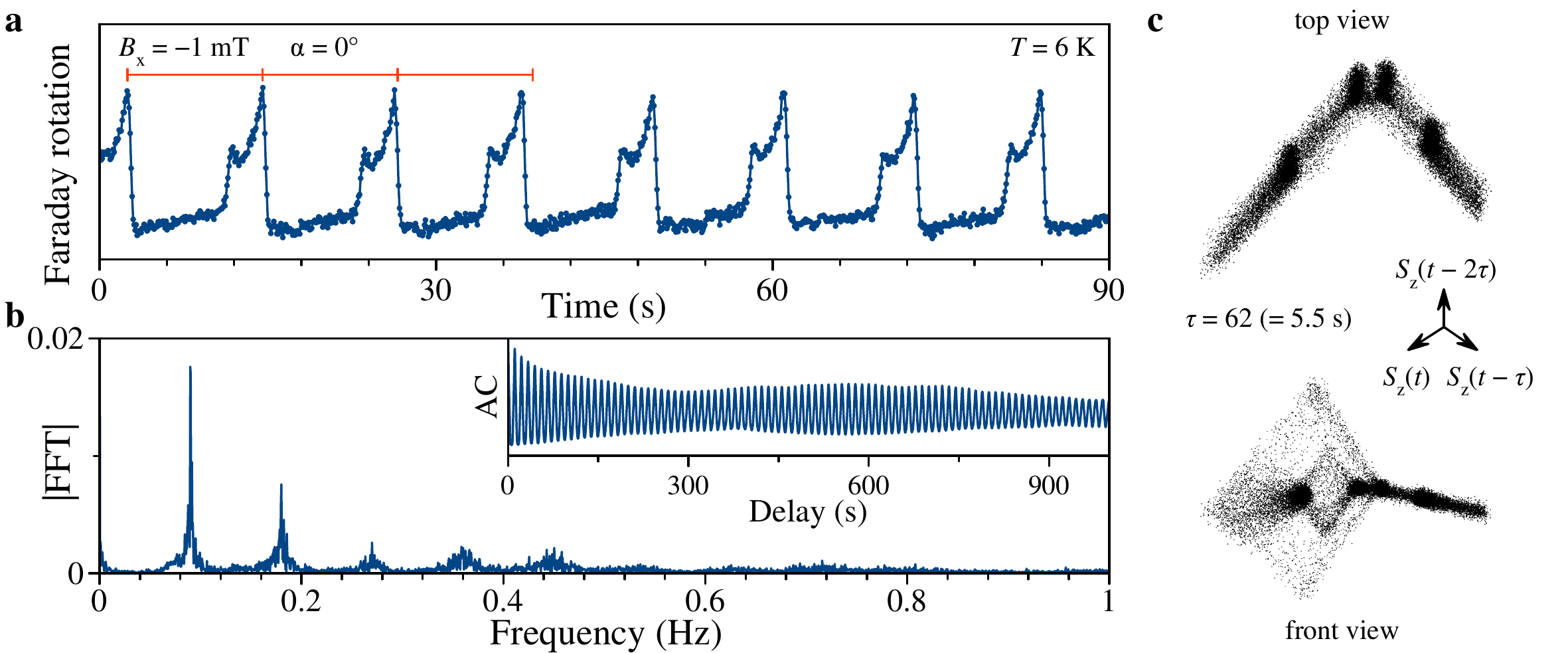}
\caption{\label{fig4} \textbf{Chaotic auto-oscillations: melting of the CTC.}
\textbf{a}, Oscillations of the electron spin polarization in the FR signal, measured in a transverse magnetic field with $B_x = -1$\,mT, $B_z = 0$\,mT ($\alpha=0^\circ$). $P_\text{pu} = 0.3$\,mW and $P_\text{pr} = 1$\,mW. The set of red time segments with equal length underlines the aperiodic behavior of the main peaks.
\textbf{b}, Fast Fourier transform and autocorrelation function vs. delay time (inset), calculated for the signal from panel \textbf{a}.
\textbf{c}, Top and front view of the phase portrait of the spin polarization auto-oscillations. $\tau=62$ time steps corresponding to 5.5\,s.}
\end{center}
\end{figure*}

Here we test the robustness of the periodic auto-oscillations with respect to variations of several experimental parameters. The Extended Data on the auto-oscillations are shown in Figs.~\ref{figS1} - \ref{figS4}, here we present the analysis of these data, applying the $0-1$ chaos test to obtain the parameter $K$, see above. The $K$-value dependencies on the photon energy of the probe, the pump and probe laser powers, the tilt and strength of the magnetic field, and the temperature of the sample are summarized in Fig.~\ref{fig3}a-d, where the black dots indicate robust parameter settings for a CTC, while the red dots give parameter settings with $K>0.99$ which are, therefore, candidates for chaotic behavior, threatening the ideal CTC behavior.

Figure~\ref{fig3}a shows the dependence of $K$ on the pump and the probe power for a fixed magnetic field orientation relative to the sample plane given by $\alpha=7^{\circ}$. The probe power does not change the FR spectra but only influences the FR signal strength. In contrast, the pump, responsible for the nuclear spin polarization, is a decisive factor: at low pump power (0.05\,mW), the efficiency of nuclear spin polarization is low, requiring about 10 minutes for saturation, see Ref.~\cite{RittmannPRB2022}. While $K=0.9926$ in this case, the signal strength is comparable to the noise level and therefore has to be considered with caution. At higher pump power (2\,mW), the signal amplitude decays due to pump-induced acceleration of the electron spin relaxation. For optimal polarization conditions, we choose 0.3\,mW pump power. 

Figure~\ref{fig3}b shows the $K$-dependence on the magnetic field angle $\alpha$ keeping $B_x=-1$\,mT fixed. At $\alpha=0^{\circ}$, the parameter $K=0.9921$, and for all other angles, $K$ is below 0.9, proving robust periodicity of the auto-oscillations. Variation of the strength of the magnetic field at fixed $\alpha=7^{\circ}$ does not give an indication for chaos, see Fig.~\ref{fig3}c. Finally, the temperature dependence of $K$ at $B_x=1$\,mT demonstrates an increase of $K$ up to $K=0.9938$ at $T=17$\,K, while below, we find CTC behavior. The condition for chaotic behavior at 17\,K, however, has to be also considered with care, as we are working here close to the edge of stability of the auto-oscillations. Even small temperature variations of about $\sim 0.1$\,K can lead to changes in the periodicity of the signal, see the Extended Data in Fig.~\ref{figS5}a. 

Summarizing all data recorded, we find that the auto-oscillations remain reliably in the strictly periodic regime, corresponding to ideal CTC behavior across wide parameter ranges. Crossing the frontiers of these stability ranges, we expect, however, chaotic behavior. We focus on the most prominent example for chaos with $\alpha=0^\circ$ (Voigt geometry). Figure~\ref{fig4}a shows a 90-second slot out of the corresponding time series of auto-oscillations, taken at $T=6\,$K for $B_x=-1\,$mT. The signal still consists of pronounced peaks with side wings at earlier delays, from first sight occurring periodically with a separation of about 12\,s. In between these features, the signal no longer vanishes but rises slowly. Taking a closer look, however, the periodicity is disturbed, as can be clearly seen from the comparison with the set of red, 12\,s long time segments in Fig.~\ref{fig4}a. 

The FFT spectrum of this time series is presented in Fig.~\ref{fig4}b and, besides the broadened peaks, shows noisy, asymmetric wings around these peaks, which have no similarity with each other and indicate a reduction of periodicity: while the oscillations are still close to periodic (see Fig.~\ref{fig4}a), taken a closer look, their period varies with time randomly. This characteristic leads to the slow decay of the autocorrelation function, see the inset of Fig.~\ref{fig4}b. Applying the time series tests, we determine the non-integer correlation dimension to be $D_2 = 1.3\pm 0.1$ and the positive maximum Lyapunov exponent to be $0.12 \pm 0.01$, corroborating the assumption of chaotic behavior. The time evolution of these auto-oscillations in phase space using delayed coordinates is shown in Fig.~\ref{fig4}c. It has a nontrivial topology with, compared to the periodic case, a tendency for a decreased phase volume that is completely filled by the data points within the limiting values. The nonlinear time series analysis for all identified cases of chaotic auto-oscillations is presented in the Extended Data, Fig.~\ref{figS6}. They all have a non-integer correlation dimension and a positive maximal Lyapunov exponent. These clear signatures for becoming chaotic and the related deviations from periodicity are a result of the melting of the CTC.

\subsection*{Interpretation of auto-oscillations}

To understand the physics behind the observations, we consider the model developed by M.~I.~D'yakonov~\textit{et al.}~\cite{DMPJETP80}, that describes periodic auto-oscillations and provides an elegant interpretation of our results.

The circularly polarized optical excitation orients electron spins, which subsequently polarize the nuclear spin system via the hyperfine interaction~\cite{OptOR_Flei_Merk}. The Overhauser field of the polarized nuclear spins $\mathbf{B}_\text{N}$ is, in general, not parallel to the average electron spin $\mathbf{S}$, so that an electron spin precesses about $\mathbf{B}_\text{N}$, causing a variation of $\mathbf{S}$. Thus, in the strongly coupled nonlinear system of electron and nuclear spins, the electron spins are responsible for the production of the Overhauser field and, vice versa, depend on its magnitude and direction. The ENSS becomes autonomous when all external parameters do not depend on time. The dissipation of angular momentum and energy in the spin system is compensated by the absorption of circularly polarized light. Then, under certain conditions, the dynamic regime of auto-oscillations may appear, which can be captured as follows: since the electron spin lifetime ($T_\text{s}$) is much shorter than the longitudinal nuclear spin relaxation time ($T_\text{N}$), the electron spin ($\mathbf{S}$) is described by the solution of the stationary Bloch equation in the sum of the external magnetic field ($\mathbf{B}_\text{ext}$) and the Overhauser field ($\mathbf{B}_\text{N}$):
\begin{equation}
\label{eq:ES}
\mathbf{S} = \mathbf{S}_0 + \frac{\mu_\text{B}gT_\text{s}}{\hbar}({\mathbf{B}_\text{ext} + \mathbf{B}_\text{N}}) \times \mathbf{S}.
\end{equation}
Here, $\mathbf{S}_0$ is the average electron spin polarization in the absence of the magnetic field, $\mu_\text{B}$ is the Bohr magneton, and $g$ is the electron $g$-factor. $\hbar/\mu_\text{B}gT_\text{s}$ is the half-width at half-maximum of the electron Hanle curve that is not influenced by dynamic nuclear polarization.

The precession of the electron spin polarization about the total magnetic field changes the Overhauser field in time according to~\cite{DMPJETP80,OptOR_Flei_Merk}:
\begin{equation}
\label{eq:NS}
\frac{d\mathbf{B}_\text{N}}{dt} = - \frac{1}{T_\text{N}}\left(\mathbf{B}_\text{N} - \hat a\mathbf{S}\right),
\end{equation}
where $\hat a$ is the second-rank tensor describing the process of dynamic nuclear polarization. The model suggests that $\hat a\mathbf{S}$ is a linear function of $\mathbf{S}$.

In the case of pure GaAs, where all lattice nuclei are located in a tetrahedral surrounding, $\hat a$ gives the contribution to the Overhauser field $B^0_\text{N}=b_\text{N}(\mathbf{S}\mathbf{B}_\text{ext})\mathbf{B}_\text{ext}/|\mathbf{B}_\text{ext}|^2$. Here $b_\text{N}$ is the parameter of the hyperfine interaction between electrons and nuclei. In this case, the values of $\mathbf{S}$ and $B^0_\text{N}$ are constant, and no auto-oscillations occur. This situation changes when the crystal symmetry is reduced.

For In$_{0.03}$Ga$_{0.97}$As, the carefully adjusted incorporation of indium replacing Ga atoms causes significant non-uniform crystal deformations (see the sketch in Fig.~\ref{fig1}a), despite the low In-content of 3\% only. The deformation affects not only the nearest neighbors of an In-atom but also more distant atoms. Accordingly, distortion magnitude and direction are highly inhomogeneous since the stress gradient decays radially with increasing distance from an inserted indium atom. The local strain leads to a quadrupole splitting of the nuclear spin levels for all nuclei with spin larger than $1/2$.

Due to the strong deformation, the spin of the $i$-th nucleus is oriented along the main local axis $\mathbf{h}_i$ of the tensor describing the quadrupole interaction rather than along the external magnetic field. The contribution of these nuclei to the total Overhauser field is $\mathbf{B}_\text{Q}=\sum_i a_i (\mathbf{S}\mathbf{h}_i)\mathbf{h}_i$, where the summation is carried out over all quadrupole perturbed nuclei within the electron localization volume around a donor~\cite{Litvyak21,Vladimirova22}. For an isotropic distribution of the axes, the field can be written as $\mathbf{B}_\text{Q}=a_\text{N}\mathbf{S}$. Therefore, $\hat{a}$ can be reduced to the simplified form~\cite{DMPJETP79}:
\begin{equation}
\label{eq:tensor-a}
\hat a\mathbf{S} = \mathbf{B}_\text{N}^0 + \mathbf{B}_\text{Q} = b_\text{N}(\mathbf{S}\mathbf{B}_\text{ext})\mathbf{B}_\text{ext}/|\mathbf{B}_\text{ext}|^2 + a_\text{N}\mathbf{S},
\end{equation}
which is the sum of the contribution to the Overhauser field from quadrupole-unperturbed nuclei ($\mathbf{B}_\text{N}^0$) and of the contribution to the Overhauser field from quadrupole-perturbed nuclei ($\mathbf{B}_\text{Q}$). The direction of the contributions and the deformation of the Hanle curve in the presence of $B_{\text{N}}^0$ as well as of $\mathbf{B}_\text{Q} + \mathbf{B}_{\text{N}}^0$ are given in the Extended Data, Fig.~\ref{figSX}.

The model allows us to simulate the periodic auto-oscillations, as shown in Fig.~\ref{fig2}c. For the calculations, we use the parameters known from the experiment: $\alpha = 10^\circ$, $B_x = -1$\,mT, and $a_\text{N} = 20$\,mT. As a fit parameter, $b_\text{N} = 21$\,mT is taken to achieve better agreement with the experiment. $T_\text{N} = 0.5$\,s is determined from the comparison of periodicity in the experimental and calculated signals. The simulated auto-oscillations reproduce the periodically repeated M-shape, in good agreement with the experimental signal. The phase space portrait is represented by a limit cycle as required for a CTC; details of the periodicity analysis are given in the Extended Data, Fig.~\ref{figSY}.

In general, periodic auto-oscillations can be observed in structures that fulfill the following conditions: (1) nuclei with a quadrupole moment are present (nuclear spin $> 1/2$), (2) the decrease of the local lattice symmetry leads to an isotropic spatial distribution of the strain-induced quadrupole splitting, and (3) the quadrupole-induced effective field $B_{\text{Q}}$ is comparably strong to the external magnetic field. 


A theoretical analysis of the coupled Eqs.~(\ref{eq:ES})-(\ref{eq:tensor-a}) was performed in Refs.~\cite{OptOR_Flei_Merk,Artemova1991}, where both the scalar and the tensor form of the nuclear fields were considered. It was shown that only limit cycles are realized, as experimentally confirmed for Al$_{0.26}$Ga$_{0.74}$As. On the other hand, Bakaleinikov~\cite{Bakaleinikov1994} showed the presence of homoclinic trajectories as a precursor for chaos, provided that the explicit form of the tensor does not correspond to a system with high symmetry -- a situation that we have achieved by our sample design. It is the crystal deformations in In$_{0.03}$Ga$_{0.97}$As facilitating the chaotic auto-oscillations.

Our experimental results indicate that chaotic auto-oscillations are observed close to the Voigt configuration. In this case, the influence of the hyperfine interaction field of the electrons on the nuclei (the Knight field) becomes important~\cite{Knight49, OptOR_Flei_Merk}. This leads to the tensor $\hat{a}=\hat{a}(S)$ becoming a nonlinear function of the electron spin $\mathbf{S}$. The general analysis of Eqs.~(\ref{eq:ES}) and~(\ref{eq:NS}) describing the ENSS is complicated and has not been carried out in full detail yet, the  importance of the Knight field for the coupled dynamics gives a hint on a further extension of the theoretical studies.

Finally, we address the controllability of the CTC period in the studied ENSS. To that end, we consider the influence of the same set of parameters on the period as in the studies of the parameter $K$. The resulting dependencies are given in Figs.~\ref{fig3}e-h, evidencing a wide tuning range of the CTC period. For example, when keeping the magnetic field orientation fixed, an increase of the field strength leads to a drop in the CTC period from about 45\,s down to almost 5\,s. Also, a change in the sample temperature leads to a similarly strong variation of the period, while the other parameter dependencies are comparatively weak. The observed variations are non-monotonic, showing that measurement of the CTC period can give further detailed insight into nonlinear systems' dynamics.

\section*{Conclusion}

The strongly correlated electron-nuclear spin system presented here provides a new dynamical many-body state in the solid state, a spin-based CTC. The hugely extended robustness and compactness compared to other CTCs allows one to explore nonlinear dynamics and seek for applications, for example, in information and metrology technology, as it may be used as a flexible frequency standard that can be controlled via sample design:
the In$_{0.03}$Ga$_{0.97}$As system with deliberately introduced symmetry reduction by deformations has facilitated the observation of auto-oscillations of the coupled ENSS across wide ranges of experimental conditions. The CTC operation is threatened by the possibility of chaotic auto-oscillations in border ranges of the parameters. 

We have confirmed the existence of this chaotic regime using analysis routines for nonlinear time series, evidencing the first such observation for semiconductor spins, possible due to the robust statistical significance of the required chaos tests. Measurement of the dynamic mode parameters shows that a strange attractor is reached in our system, as evidenced by a non-integer correlation dimensionality and a positive maximum Lyapunov exponent. The description of deterministic chaos in the ENSS still remains a challenge asking for an extended mathematical analysis. For further progress on the experimental side, one may consider active modulation of the external CTC parameters in time, thereby getting active control of the periodic auto-oscillations. 


\bibliographystyle{naturemag}


\subsection*{Acknowledgements}
The authors are thankful to D.~R. Yakovlev for fruitful discussions. We acknowledge financial support by the Deutsche Forschungsgemeinschaft in the frame of the International Collaborative Research Centre TRR160 (Project A1). A.G. and M.B. acknowledge support by the BMBF project QR.X (Contract No.16KISQ011). The Resource Center "Nanophotonics" of Saint-Petersburg State University provided the epilayer sample.

\subsection*{Author contributions}
V.L.K. and A.G. conceived the experiment. A.N.K., P.S.S., and A.G. built the experimental apparatus and performed the measurements. A.N.K., N.E.K., P.S.S., V.L.K., and A.G. analyzed the data. N.E.K. and V.L.K. provided the theoretical description. All authors contributed to the interpretation of the data. N.E.K., V.L.K., and A.G. wrote the manuscript in close consultation with M.B., supervising the project.


\newpage

\setcounter{equation}{0}
\setcounter{figure}{0}
\renewcommand{\figurename}{\textbf{Extended Data Figure}}

\section*{Methods}

\textbf{Sample.}
The sample under study is a $10\,\mu$m thick InGaAs epilayer with 3\% indium content. It was grown by molecular-beam epitaxy on a GaAs substrate and is homogeneously doped by Si atoms. At low temperatures ($T\approx 2\,$K), the donor density is $4.2\times 10^{15}$\,cm$^{-3}$, thoroughly determined in Ref.~\cite{RittmannPRB2022}. An important effect of the indium incorporation is the strongly enhanced quadrupole splitting, leading to local nuclear fields of $B_Q=20\,$mT~\cite{RittmannPRB2022}. This value is an order of magnitude larger than in low-stressed GaAs structures~\cite{VladimirovaPRB22} and two orders of magnitude larger than in GaAs epilayers with a quadrupole-free local field, given purely by the dipole-dipole interaction of the nuclei~\cite{VladimirovaPRB18}. 

\begin{figure*}[b!]
\begin{center}
\includegraphics[width = 16cm]{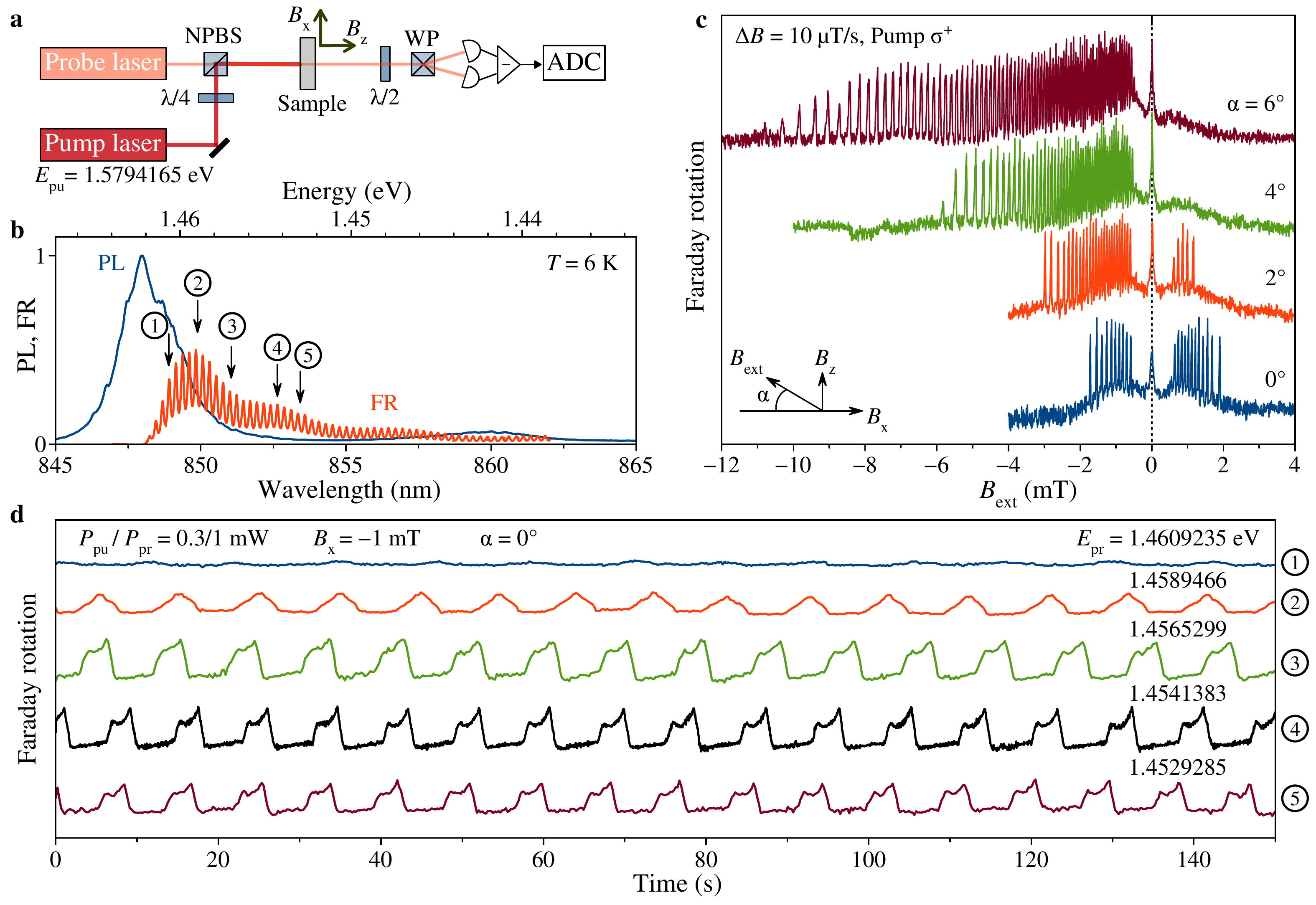}
\caption{\label{figS1} \textbf{Setup and auto-oscillations.} \textbf{a}, Schematic of the experimental setup with the pump and probe laser beams. See the text for a detailed description of the components. \textbf{b}, Normalized photoluminescence (PL) (blue curve) and Faraday rotation (red curve) of the sample. Black arrows mark the photon energies of the probe laser for the time traces in panel \textbf{d}. \textbf{c}, FR vs. magnetic field at different angles $\alpha$ of the external magnetic field relative to the sample plane for constant $\sigma^+$ pump excitation. $E_\text{pu}=1.5794165\,$eV, $E_\text{pr} = 1.4541383\,$eV. Pump/Probe powers $P_\text{pu}/P_\text{pr}=0.3/1\,$mW. Magnetic field sweeping rate $\Delta B= 10\,\mu$T/s. The sketch at the bottom left shows the direction of the magnetic field and its components. \textbf{d}, Auto-oscillations measured at different photon energies of the probe laser for $\alpha=0^{\circ}$ at $B_x=-1\,$mT. $T=6$\,K.}
\end{center}
\end{figure*}

\textbf{Experimental details.}
Extended Data Fig.~\ref{figS1}a schematically shows the experimental setup. We use a continuous wave (CW) diode laser emitting at $E_\text{pu}=1.579\,$eV photon energy (785\,nm) as a pump source to optically excite an electron spin polarization. A Ti:Sapphire CW ring laser with tunable emission energy is used as a probe source to measure the spin polarization of the resident electrons using Faraday rotation. This laser is stabilized in wavelength down to 0.001\,nm using a high-precision wavemeter (HighFinesse WS6-600) and a piezo controller for the laser cavity length with proportional-integral-differential (PID) control. Before hitting the sample, the circularly polarized pump and the linearly polarized probe laser beams are combined using a non-polarizing beam splitter (NPBS). Due to the high energy non-resonant excitation by the pump (above the GaAs band gap), the pump-laser beam is absorbed by the GaAs substrate and does not enter the detection channel. This allows us to use a collinear scheme in which pump and probe take the same optical path and hit the sample along the $z$ axis. The probe laser is tuned below the GaAs absorption edge so that it is transmitted through the sample. Both beams are focused onto the sample surface to spots with diameters of 13\,$\mu$m for the probe and 40\,$\mu$m for the pump. The emission powers of both lasers are stabilized by laser power controllers (noise eaters) down to 0.03\% for the probe and 0.05\% for the pump within the DC$-4$\,kHz range. The InGaAs epilayer is mounted in a helium flow cryostat in the center of two orthogonal pairs of electromagnet coils, generating the magnetic field components $B_\text{x}$ and $B_\text{z}$. Using the temperature controller for the heater voltage and the helium flow, one can set a temperature in the range of 5\,K up to 300\,K. At low temperatures ($T\approx 5$\,K) the precision of temperature stabilization reaches the standard deviation of $\sigma = 0.002$\,K using a PID controller, while in the range around $T\approx 17\,$K, the stability is $\sigma = 0.017$\,K. After transmission through the sample, the induced FR of the probe beam is detected using a half-wave plate followed by a Wollaston prism and a balanced photoreceiver. The differential signal is stored after digitization.

Extended Data Fig.~\ref{figS1}b demonstrates the photoluminescence (PL) of the InGaAs epilayer with the emission maximum at 1.462\,eV ($\sim 848$\,nm), the blue curve. The red curve shows the Faraday rotation signal at zero magnetic field, measured by scanning the photon energy of the probe laser with constant incident power. The sample was polished from the back side to allow for transmission measurements and covered from both sides with antireflection coatings to reduce etalon effects. Nevertheless, some residual effects are observed in the FR measurements as seen by the oscillating red curve in Extended Data Fig.~\ref{figS1}b. The same curve is shown in the main text in Fig.~1a with smoothed oscillations to avoid confusion from their similarity with the FR auto-oscillations. 

Extended Data Fig.~\ref{figS1}c shows Faraday rotation signals recorded in magnetic field scans at different angles $\alpha$ of the field vector relative to the sample plane, see the sketch in the panel. For each scan, $\alpha$ is kept constant while the magnetic field is swept at a rate of $10\,\mu$T/s. A slower sweep rate increases the frequency of the oscillations but does not increase the field range in which they appear. The curves do not depend on the field scan direction. The blue bottom curve demonstrates the case when the magnetic field is applied in the sample plane transversely to the excitation $z$ axis. The curve has a symmetric shape with a narrow peak at zero magnetic field and two broad peaks located symmetrically to zero field. These broad peaks demonstrate the points of compensation of the internal effective nuclear field (Overhauser field of about 1\,mT strength) seen by the electrons by the external magnetic field, compare with Extended Data Fig.~\ref{figSX}a. On top of the broad side peaks, one can observe an additional modulation of the signal that disappears at higher fields. 

Several time traces measured at different probe energies with the transverse magnetic field kept constant at $B_x=-1$\,mT ($\alpha=0^{\circ}$), are demonstrated in the Extended Data Fig.~\ref{figS1}d. We expect that we measure the same localized electron states in every case while the FR has a smooth probe energy dependence decaying when moving away from the resonance. For the experiments in the main text, we measure at the probe energy of $E_\text{pr}=1.454\,$eV ($852.63\,$nm), where the FR signal demonstrates a local maximum in combination with weak absorption (Extended Data Fig.~\ref{figS1}b) and shows  auto-oscillations of strong amplitude. 

We do not use any modulation of the employed lasers or the external parameters and no lock-in amplifier in the experiments. All signals are measured as DC signals with an integration time of 4.45\,ms. The electro-optical modulator for the pump beam and the lock-in amplifier is only used once to measure the Hanle curve for alternating pump laser helicity, in order to avoid nuclear spin polarization.

\begin{figure}[t!]
\begin{center}
\includegraphics[width = 8cm]{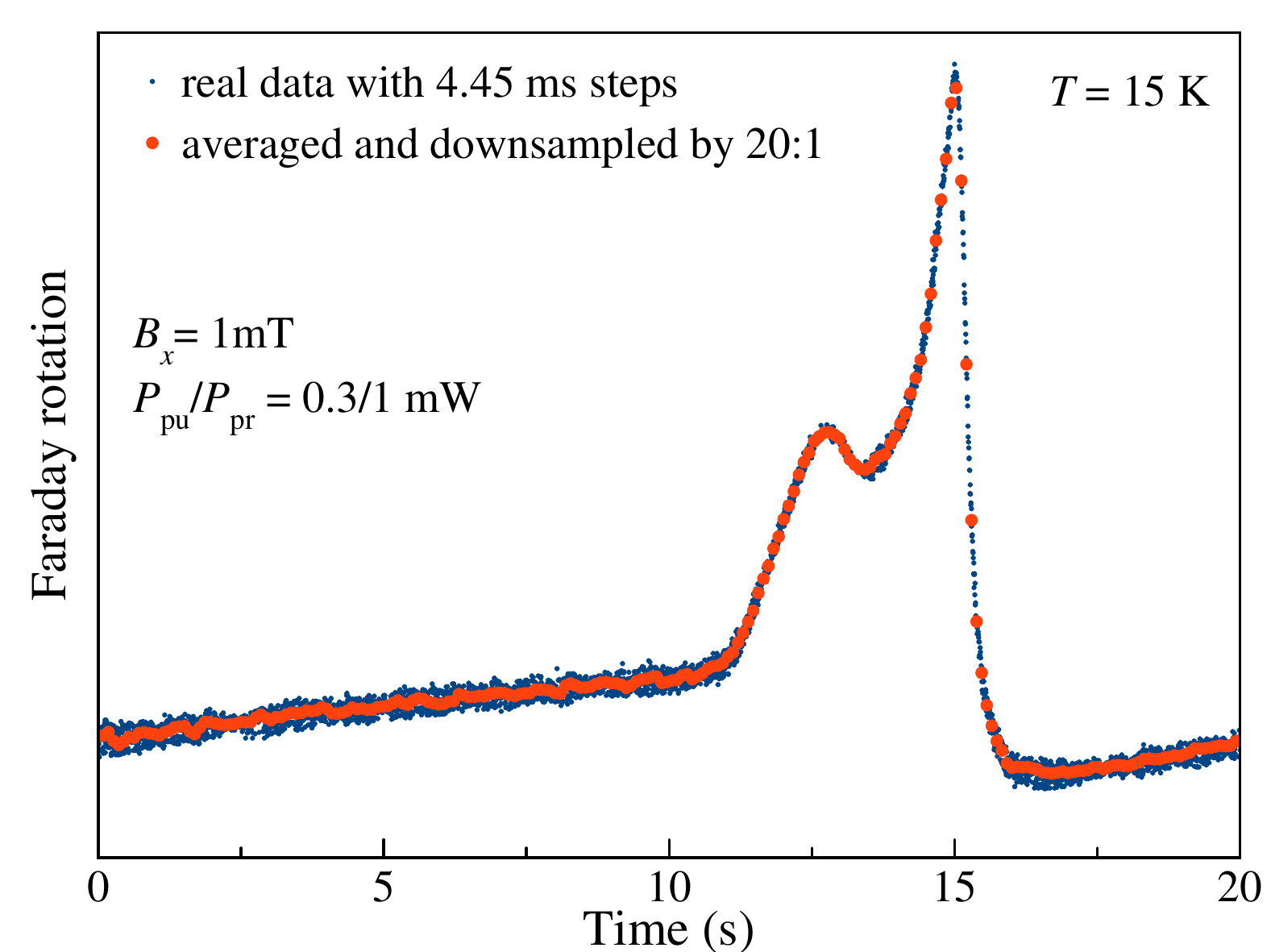}
\caption{\label{figS2} \textbf{Smoothing and down-sampling.}
Fraction of an exemplary raw data signal (blue points) combined with demonstration of the smoothing and down-sampling effects (red points). 
}
\end{center}
\end{figure}

For further signal processing, we used a Savitzky-Golay filter~\cite{SGF} with a second-order polynomial using 21 window points to smooth the data and down-sampled the signal by a factor of 20, as shown in Extended Data Fig.~\ref{figS2}. The usual signal accumulation time for time traces was 40 minutes, giving us about 30,000 time points after the down-sampling to have sufficient statistics for the nonlinear signal processing.

\begin{figure*}[b!]
\begin{center}
\includegraphics[width = 18cm]{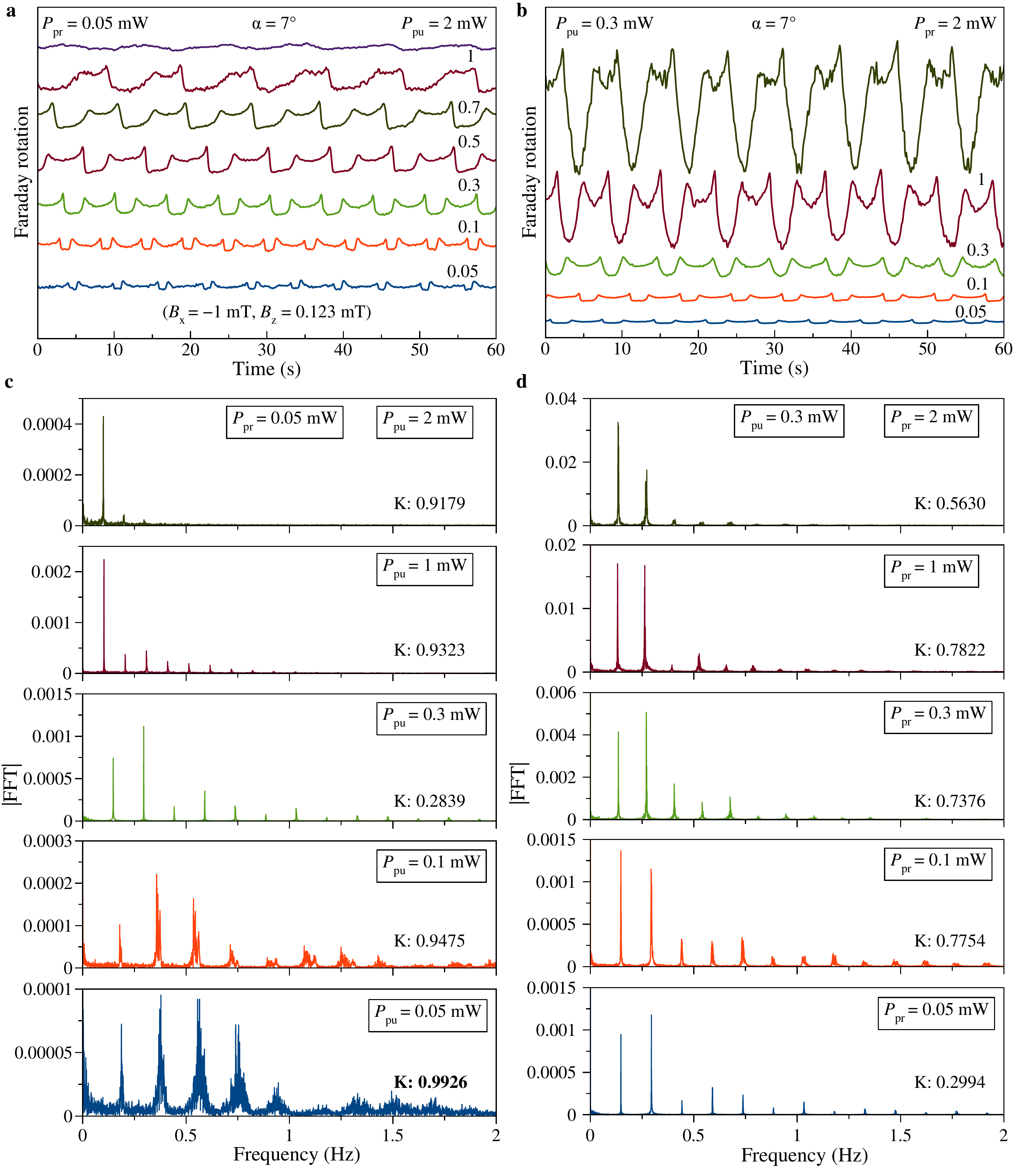}
\caption{\label{figS3} \textbf{Laser power dependence.}
Time series measured for various pump powers at fixed $P_\text{pr} = 0.05$\,mW, \textbf{a}, and for various probe powers at fixed $P_\text{pu} = 0.3$\,mW, \textbf{b}. $E_\text{pr} = 1.4541383$\,eV ($\lambda_\text{pr}=852.63\,$nm), $E_\text{pu} = 1.5794165$\,eV ($\lambda_\text{pu}=785\,$nm). $\alpha = 7^{\circ}$ and $T = 6$\,K. \textbf{c} and \textbf{d}, FFT spectra of the auto-oscillations in \textbf{a} and \textbf{b}. $K$ is the parameter resulting from the modified $0-1$ test for chaos.}
\end{center}
\end{figure*}

\textbf{Pump- and probe-power dependence.} 
Extended Data Figs.~\ref{figS3}a and~\ref{figS3}b demonstrate the first minute of the time series for varying pump and probe powers. The corresponding Fourier spectra are presented in Extended Data Fig.~\ref{figS3}c and allow one to detect the variation of the frequency distributions corresponding to CTC structural analyses. As known from Ref.~\cite{RittmannPRB2022}, the nuclear spin polarization saturates within 5 minutes at 0.3\,mW pump power for the same configuration. Therefore, we wait for five minutes before starting to record a new time series so that any transient effects have settled.

Extended Data Figure~\ref{figS3}b shows the effect of the probe power on the FR signal for 0.3\,mW pump power. We observe a linear increase in the signal amplitude with probe power. Further, the probe power weakly affects the frequency distribution, as demonstrated by the Fourier spectra in Extended Data Fig.~\ref{figS3}d. Note that the relative weight of the Fourier peaks varies, however, indicating changes in the crystal unit cell, but not the crystal period. In the following, we use the probe power of 1\,mW to obtain the best signal-to-noise ratio. 

We have applied the Chaos Decision Tree Algorithm presented in Ref.~\cite{Toker2020} to all data sets for the nonlinear time series analysis. The corresponding results are presented in the main text, Fig.~3. As we know that the system's evolution is given by a nonlinear set of equations, we use the Cycle Shuffled Surrogates to test our data for the dynamical correlation between the cycles~\cite{THEILER1995335,LANCASTER20181}. Further parameters of the test are set to the default values: we use the Schreiber algorithm for the data denoising~\cite{SchreiberPRE472401}, check for oversampling, and test for chaos without giving a specific value for the cutoff $K$~\cite{Gottwald01test}. 

\begin{figure*}[t!]
\begin{center}
\includegraphics[width = 18cm]{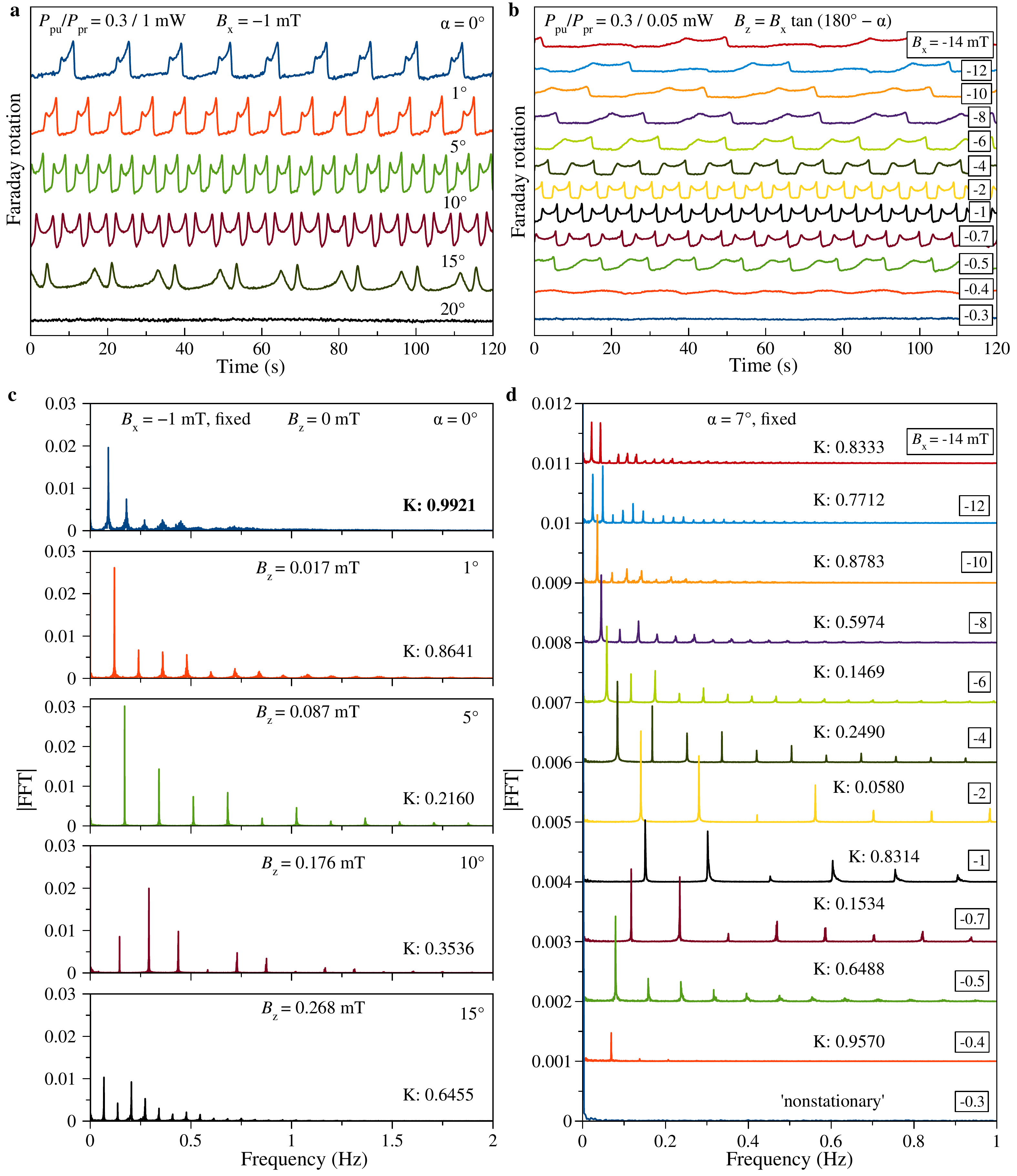}
\caption{\label{figS4} \textbf{Magnetic field dependence.} \textbf{a}, Time series measured at different angles $\alpha$ for fixed $B_\text{x}=-1$\,mT. $P_\text{pu} = 0.3$\,mW, $P_\text{pr} = 1$\,mW. \textbf{b}, Time series measured for fixed angle $\alpha=7^{\circ}$ and at different strength of the magnetic field. Both $B_x$ and $B_z$ are varied. $P_\text{pu} = 0.3$\,mW, $P_\text{pr} = 0.05$\,mW. \textbf{c} and \textbf{d} are the corresponding Fourier transform spectra. $T =6$\,K.}
\end{center}
\end{figure*}

\textbf{Magnetic field strength and angle dependence.}
Extended Data Fig.~\ref{figS4} demonstrates the dependence of the time series evolution on the direction and strength of the external magnetic field. In the main text, Fig.~2 demonstrates the exemplary data set for the periodic case with $\text{K}=0.3536$, measured at $\alpha = 10^{\circ}$ and $B_x=-1\,$mT as well as $B_z=0.176\,$mT.

\begin{figure*}[t!]
\begin{center}
\includegraphics[width = 18cm]{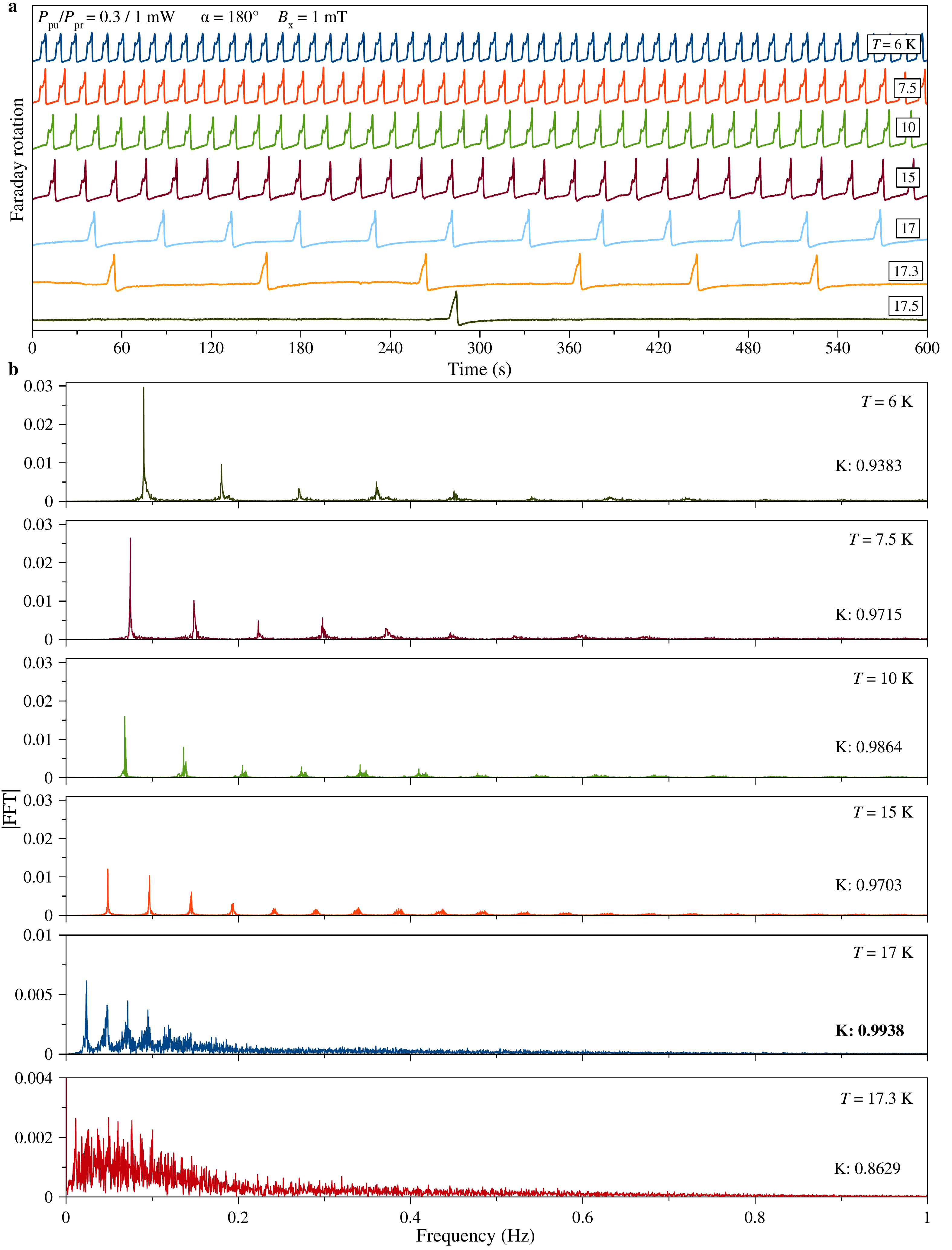}
\caption{\label{figS5} \textbf{Temperature dependence.}
\textbf{a}, Time series measured for $\alpha=180^{\circ}$ and $B_x=1\,$mT at different temperatures. $P_\text{pu} = 0.3$\,mW,  $P_\text{pr} = 1$\,mW. \textbf{b}, corresponding Fourier spectra.}
\end{center}
\end{figure*}

\textbf{Temperature dependence.}
Extended Data Fig.~\ref{figS5}a shows the first 10 minutes of time series measured at different temperatures, with Extended Data Fig.~\ref{figS5}b demonstrating their Fourier spectra. A prominent feature in the data is the reduction of the system's frequencies with rising temperature from 6\,K up to 17\,K, which suggests a variation of the nuclear spin relaxation time $T_\text{N}$. In Ref.~\cite{RittmannPRB2022}, the temperature dependence of the integral spin noise is constant up to 20\,K, proving charge stability of the donor-bound electrons. The electron spin relaxation time starts to reduce above $T = 10\,$K, for example, by a factor of two at about 20\,K. The evaluated $K$-values indicate chaotic behavior around $T=17\,$K with $\text{K}=0.9938$.

\textbf{Correlation dimension and maximal Lyapunov exponent.}
Here we provide additional measures and discriminating statistics established for testing whether a system is chaotic. These include the correlation dimension ($D_2$)~\cite{Grassberger1983} and the maximal Lyapunov exponent (LE)~\cite{ROSENSTEIN1993117}. In our case, these are calculated using the TISEAN software package for nonlinear time series analysis~\cite{Hegger1999}.

\begin{figure*}[t!]
\begin{center}
\includegraphics[width = 18cm]{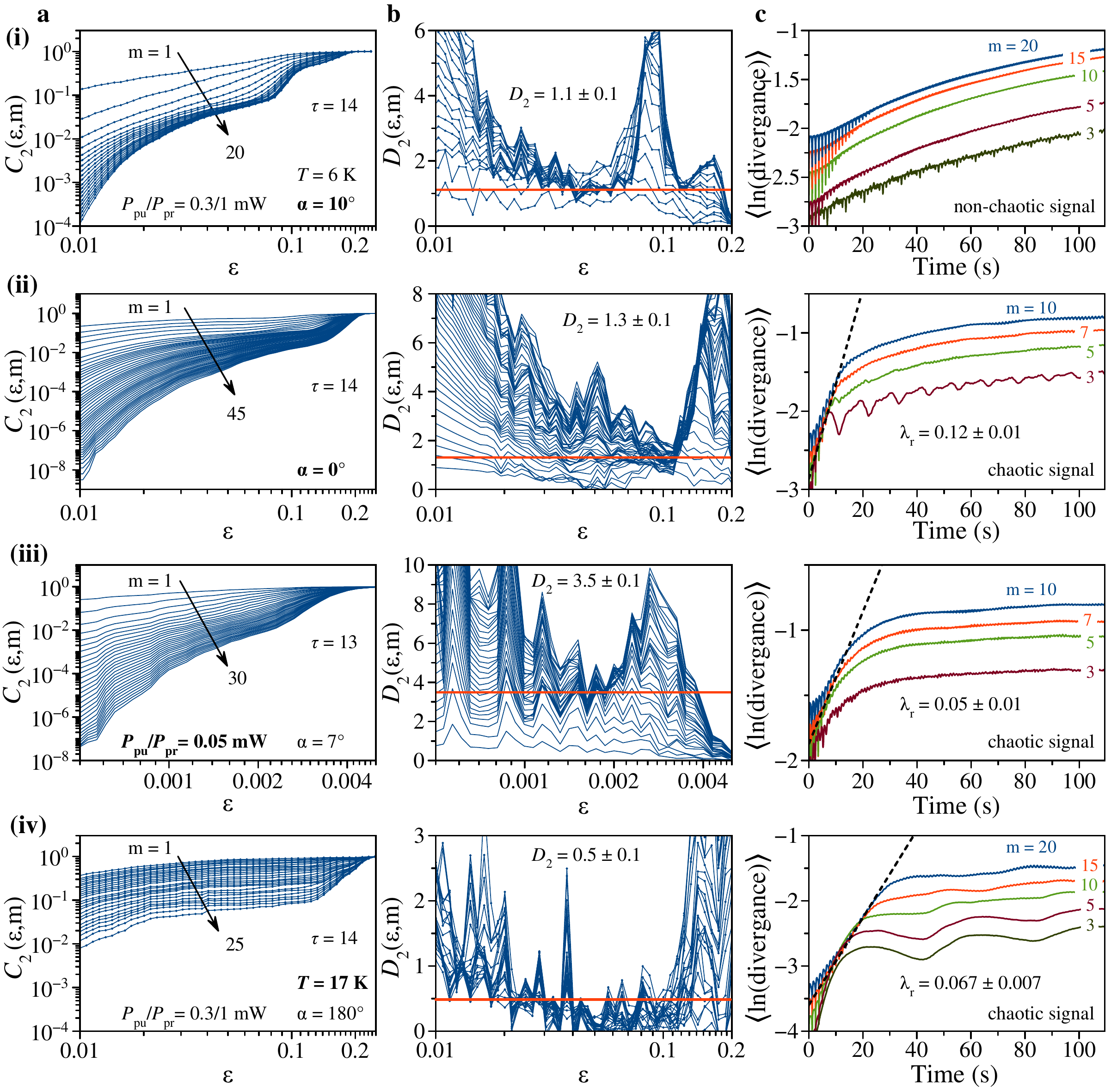}
\caption{\label{figS6} \textbf{TIme SEries ANalysis.} 
The data are sorted in rows with similar analysis. The first row \textbf{(i)} presents the data for $\alpha=10^{\circ}$, $T=6\,$K, and $P_\text{pu}/P_\text{pr}=0.3/1\,$mW. \textbf{a}, Log-Log plot of $C_2(\varepsilon)$ vs. $\varepsilon$ for increasing embedding dimension $m$ (top to bottom). \textbf{b}, Plot of the slope in panel \textbf{a} as function of $log(\varepsilon)$, which converges to a constant value for a range of $\varepsilon$, shown by the red line. \textbf{c}, Calculation of the average logarithm of the divergence versus time for increasing embedding dimensions. The second row \textbf{(i)} presents the data for $\alpha=0^{\circ}$, $T=6\,$K, and $P_\text{pu}/P_\text{pr}=0.3/1\,$mW. The black dashed line in the last panel for determining the LE marks the linear scaling region. The third row \textbf{(iii)} presents the data for $\alpha=7^{\circ}$, $T=6\,$K, and $P_\text{pu}/P_\text{pr}=0.05\,$mW. The fourth row \textbf{(iv)} presents the data for $\alpha=180^{\circ}$, $T=17\,$K, and $P_\text{pu}/P_\text{pr}=0.3/1\,$mW.}
\end{center}
\end{figure*}

The first row (i) of Extended Data Fig.~\ref{figS6} presents an analysis of the data in the main text, Fig.~2. Extended Data Fig.~\ref{figS6}a demonstrates the log-log plot of the correlation sum $C_2$ vs. increasing size of the threshold radius $\varepsilon$ for an increasing number of embedding dimensions $m$. The embedding time delay $\tau=14$ indicated in the same panel is evaluated by calculating the mutual information using the TISEAN package~\cite{FraserMutual}. 14 dimensionless time steps correspond to 14\,(steps)*0.0889\,(s/step)$=1.2446$\,s. Extended Data Fig.~\ref{figS6}b shows the slope of the curves from panel~\ref{figS6}a for all embedding dimensions. The linear range of converged curves is observed around the $\varepsilon$ range of 0.04 to 0.07, leading to $D_2=1.1 \pm 0.1$, shown by the red line. This value is close to an integer, indicating the periodic character of the time series. Extended Data Fig.~\ref{figS6}c demonstrates calculations of the maximal LE using the Rosenstein algorithm~\cite{ROSENSTEIN1993117} for different embedding dimensions. The absence of a linear scaling region gives no convincing result. Taking into account all other metrics, like the integer correlation dimension and the $K$-value of 0.3536 from the chaos test, we can conclude that the data indicate non-chaotic behavior.

The second row (ii) of Extended Data Fig.~\ref{figS6} presents the analysis of the second data set presented in the main text, Fig.~3. Following the same procedure, we discover that the correlation dimension has the non-integer value of $D_2= 1.3$. Furthermore, Rosenstein's algorithm for the maximal LE exhibits a clear range of a linear increase with an identical slope for the rising embedding dimension $m$. This slope can be taken as an estimate of the maximal exponent $\lambda_\text{r}=0.12$. Taking into account the $K=0.9921$, the non-integer value of $D_2=1.3$, and the positive value of the maximal LE, we conclude that the time series for $B_x=-1\,$mT and $\alpha=0^{\circ}$ is chaotic.

The third row (iii) of Extended Data Fig.~\ref{figS6} presents the data analysis from the power dependence. Here $D_2= 3.5$ and $\lambda_\text{r}=0.05$. Considering the $K=0.9926$, the non-integer value of $D_2=3.5$, and the positive value of the maximal LE we conclude that the time series for $P_\text{pu}/P_\text{pr}=0.05\,$mW and $\alpha=7^{\circ}$ is chaotic as well. The fourth row (iv) of Extended Data Fig.~\ref{figS6} demonstrates the value of $D_2= 0.5$ and $\lambda_\text{r}=0.067$. Based on the $\text{K}=0.9938$, the non-integer value of $D_2$, and the positive value of the maximal LE, the time series for $T=17\,$K and $\alpha=180^{\circ}$ is chaotic.

\begin{figure*}[h]
\begin{center}
\includegraphics[width = 16cm]{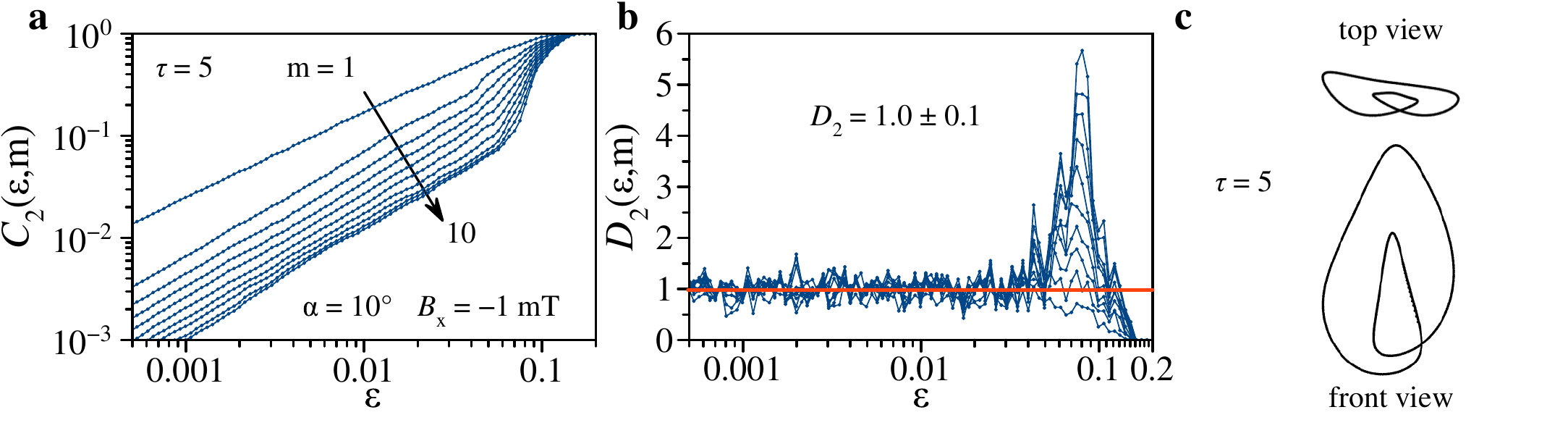}
\caption{\label{figSY} \textbf{Nonlinear time series analysis for the simulated data.} \textbf{a}, correlation sum calculated for the simulated data set for the time delay $\tau=5$ and for different embedding dimensions $m$. \textbf{b}, corresponding slopes with the evident horizontal range having a slope value of $D_2=1$. \textbf{c}, phase portrait calculated for the time delay $\tau=5$.}
\end{center}
\end{figure*}

\begin{figure*}[t!]
\begin{center}
\includegraphics[width=16cm]{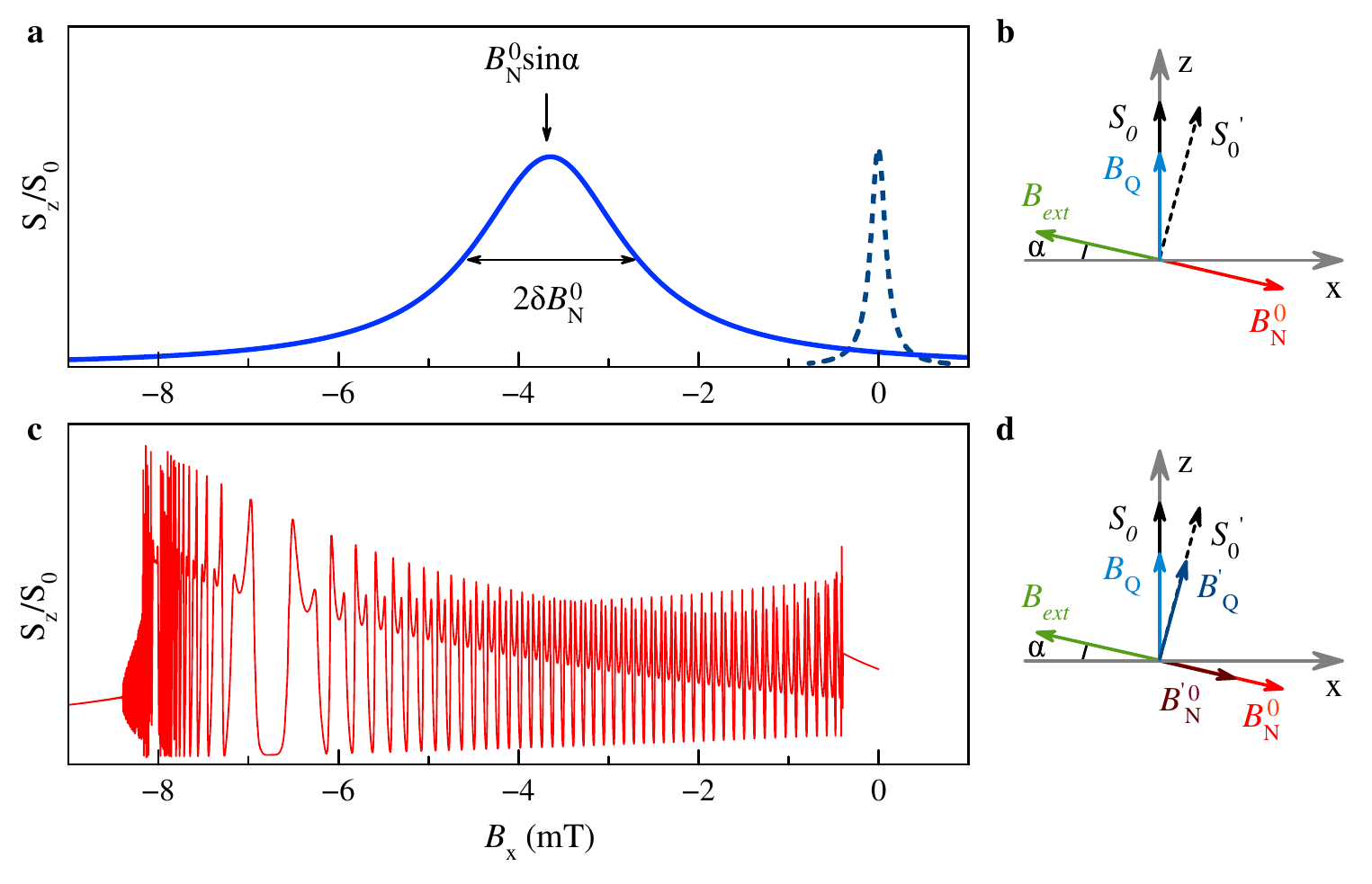}
\caption{\label{figSX} \textbf{Electron spin polarization in the transverse magnetic field in the case of dynamic nuclear spin polarization.}
\textbf{a}, Hanle curve for electrons (blue dotted curve), and FR amplitude for electrons experiencing a stationary nuclear field (solid blue curve) as a function of the transverse magnetic field. \textbf{b}, Scheme of formation of the stationary Overhauser field $\mathbf{B}_\text{N}$ along the external magnetic field $\mathbf{B}$ and the effective magnetic field from quadrupole perturbed arsenic nuclei $B_{\text{Q}}$ along the electron spin polarization $\mathbf{S}$. \textbf{c}, FR amplitude for electrons experiencing the Overhauser field as a function of the transverse magnetic field. \textbf{d}, Scheme of formation of the Overhauser field. The dashed index indicates the deviation of the spin and the reorientation of the Overhauser field from the original values.}
\end{center}
\end{figure*}

To finalize our analysis, we have evaluated the $C_2$ and $D_2$ values for the simulated data presented for the periodic case in Fig.\,2 of the main text, see Extended Data Figs.~\ref{figSY}a and~\ref{figSY}b. The $D_2=1$ coincides well with the $D_2$ estimated from the experimental time series presented in Extended Data Fig.~\ref{figS6}b. Extended Data Figure~\ref{figSY}c additionally shows the phase portrait, which compares well with the one shown in Fig.\,2 of the main text.

\textbf{Model of auto-oscillations.}
The general analysis of Eqs.~\ref{eq:ES} and~\ref{eq:NS} for the ENSS is complicated and has not yet been carried out in all detail. For the simplest case of an isotropic spatial distribution of the principal axes of the quadrupole interaction tensor, the contribution of the quadrupole perturbed nuclei, when averaged over all nuclei in the In neighborhood, is reduced to the isotropic form $a_\text{N} \cdot \delta_{\alpha\beta}$. Then, the averaged $\mathbf{B}_{\text{Q}} = a_\text{N}\mathbf{S}$ is aligned along the electron spin ($\mathbf{S}$), as shown in Extended Data Fig.~\ref{figSX}b by the blue arrow. 

The unperturbed nuclear spin sublevels create the Overhauser field $\mathbf{B}_\text{N}^0$ as in pure, unstressed GaAs. $\mathbf{B}_\text{N}^0$ is aligned along the external magnetic field, as shown by the red arrow in Extended Data Fig.~\ref{figSX}b, and compensates for the Zeeman splitting of the electrons in the external magnetic field so that the FR maximum in the transverse field is shifted from $B_x = 0$ (see the blue dotted line in Extended Data Fig.~\ref{figSX}a) to the position $B_x = B_\text{N}^0\sin\alpha$ (shown by the solid blue line). We assume that $B_\text{N}^0$ is highly inhomogeneous, causing the broadening of the curve described by the half-width of half maximum $\delta B_\text{N}^0$. We use the parameter $\delta B_\text{N}^0 = 1$\,mT for a better match with the experiment.

The joint action of $B_\text{N}^0$ and $B_{\text{Q}}$, which combine to the Overhauser field, can lead to instability of the ENSS, as can be understood in the following way: The electron spin is oriented by the circularly polarized pump along its propagation direction $\mathbf{k}$ ($\mathbf{S}_0$ in Extended Data Fig.~\ref{figSX}d). Subsequently, the electron spin polarization averages out due to Larmor precession around the total magnetic field (external field plus Overhauser field) and deviates from the direction determined by the pump light ($\mathbf{S'}$ in Extended Data Fig.~\ref{figSX}d). This new spin orientation leads to a change of $\mathbf{B}_\text{N}^0$ and to a reorientation of the average $\mathbf{B}_\text{Q}$, as both depend on the spin. The adjusted values of $\mathbf{B}_\text{N}^0$ and $\mathbf{B}_\text{Q}$ are shown in Extended Data Fig.~\ref{figSX}d as ${\mathbf{B}_\text{N}^0}'$ and $\mathbf{B}_\text{Q}'$. This steadily ongoing process results in the further evolution of the average electron spin, leading to a new configuration of the Overhauser field, and so on. Therefore, continuous auto-oscillations of the spin polarization are created. The longitudinal nuclear spin relaxation time gives the process's periodicity. Following the calculations from Ref.~\cite{DMPJETP80}, instability occurs in the range of external magnetic fields, where $\tilde{a}_\text{N}\tilde{b}_\text{N}\cos^2\alpha > 2$, where $\tilde{a}_\text{N} = {a}_\text{N}\mathbf{S}_0/\delta B_\text{N}^0$ and $\tilde{b}_\text{N} = {b}_\text{N}\mathbf{S}_0/\delta B_\text{N}^0$~\cite{Kalevich1986,Artemova1991}. In this case, the steady-state spin polarization turns out to be unstable~\cite{DMPJETP80}, as shown in Extended Data Fig.~\ref{figSX}c, resulting in a limit cycle.

It is worth applying the gained knowledge to self-assembled quantum dot (QD) structures, which are currently intensively studied for quantum information applications~\cite{Gershoni16,Atatuere19,Rastelli21}. The intrinsic, strain-induced quadrupole splitting in these structures leads to $B_\text{Q}$ values on the order of 300\,mT~\cite{Maletinsky2009}, thereby fulfilling two out of three conditions mentioned in the paper, namely (1) and (3). Condition (2), however, is not matched in these QDs. As shown in Ref.~\cite{Sokolov16}, the strain gradient in InGaAs QDs is mostly oriented along the growth axis, parallel to the [001] crystal direction, with a deviation of maximally 8 degrees, leading to ENSS stability~\cite{Dzhioev07}.


\end{document}